\documentclass[onecolumn,showpacs,10pt]{revtex4}

\usepackage{graphicx}% Include figure files
\usepackage{dcolumn}% Align table columns on decimal point
\usepackage{bm}% bold math
\usepackage{amsmath}
\usepackage{amssymb}
\usepackage{graphics}
\newcommand{\mathsym}[1]{{}}

\input epsf

\begin{document}

\title{\Large{Reconstruction of Einstein-Aether Gravity from other Modified Gravity Models }}

\author{\bf Chayan Ranjit$^1$\footnote{chayanranjit@gmail.com}, and Ujjal
Debnath$^2$\footnote{ujjaldebnath@yahoo.com}}
\affiliation{$^{1}$Department of Mathematics, Seacom Engineering College, Howrah - 711 302, India.\\
$^{2}${Department of Mathematics, Indian Institute of Engineering
Science and Technology, Shibpur, Howrah-711 103, India.} }

\date{\today}

\begin{abstract}
We briefly describe the modified Friedmann equations for
Einstein-Aether gravity theory and we find the effective density
and pressure. The purpose of our present work is to reconstruction
of Einstein-Aether Gravity from other modified gravities like
$f(T)$, $f(R)$, $f(G)$, $f(R,T)$ and $f(R,G)$ and check its
viability. The scale factor is chosen in power law form. The free
function $F(K)$ for Einstein-Aether gravity (where $K$ is
proportional to $H^{2}$) have been found in terms for $K$ by the
correspondence between Einstein-Aether gravity and other modified
gravities and the nature of $F(K)$ vs $K$ have been shown
graphically for every cases. Finally, we analyzed the stability of
each reconstructed Einstein-Aether gravity model.
\end{abstract}

\pacs{04.50.Kd, 95.36.+x, 98.80.Cq, 98.80.-k}

\maketitle

\section{\normalsize\bf{Introduction}}
Recently different observational study of the luminosity type Ia
supernovae (SNIa)
\cite{Gold2009,Nolta2009,Bachall1999,Perlmutter1999} and Cosmic
Microwave Background (CMB) \cite{Perlmutter1998,Riess,Riess1,
Bennet,Sperge} radiation strongly indicate that our universe is
currently expanding with an acceleration. There are many
candidates which are considered as the main responsible for this
expanding scenario. Dark energy (DE)
\cite{Sahni2006,Seikel,Clarkson,Liu} is one of them. There are
different exotic types of dark energy with negative pressure. It
can be separated by depending on their equation of state (EOS).
When EOS is $-1<\omega<-1/3$, they are called quintessence type DE
and when $\omega<-1$, they are called phantom type DE. There are
some other type of dark energies model which can cross the phantom
divide $\omega=-1$ from both sides and are called quintom type DE.
Another idea to describe such acceleration scenarios is understood
by imposing a concept of modification of gravity for an
alternative candidate of dark energy. This type of model provides
very natural gravitational alternative for exotic matter. This
type of gravity models are predicted by string/M-theory. Modify
Gravity models describe the phantom or non-phantom or quintom
phase of the universe by without introducing negative kinetic term
of dark energies. It gives a natural description about the early
time inflation then transition from deceleration to late time
acceleration. As a rival of General Relativity, some form of the
theory of modified gravity are now very important for present
research scenario while there are some rigid constraints in modify
gravity theory. Cosmic acceleration of Dark energy models can also
be explained by modifying in Ricci curvature $R$ of Lagrangian.
Several form of Ricci curvature have been considered by the
different researcher \cite{Vollick,Briscese,
Carroll,Abdalla,Linder} to explain the accelerating phase of the
Universe. Remarkable results are obtained by some work of Nojiri
et al \cite{Nojiri2003,Nojiri2007} in which they considered the
modified Lagrangian as $R+ R^m + R^{-m}$ and as a result they
obtained an inflation at an early stage and also a late time of
accelerated expansion. These result lead to open new window of
gravity dependent research methodology. The most popular modified
models of gravity include $f(R)$ gravity (where $R$ represent the
Ricci Scalar Curvature) \cite{Nojiri2007a,Li,Nojiri2006}, $f(T)$
gravity (where $T$ is the torsion scalar)
\cite{Li2,Myrzakulov,Wu}, $f(G)$ gravity (where $G$ represents the
Gauss-Bonnet invariant) \cite{Rastkar,Nojiri2005}, $f(R,T)$
gravity \cite{Harko2011,Jamil2012}, $f(R,G)$ gravity
\cite{Bamba2010,Myrzakulov2013} and many more.\\

Recently, reconstruction between different dark energy models
become a very popular scenario among the researchers in
Cosmological study. In 2008, Setare et al. \cite{Setare2008}
investigated the cosmological implications of the correspondence
of Holographic dark energy model and Gauss-Bonnet dark energy
model. This leads to a explanation of the accelerated expansion of
the universe by imposing specific constraints. Liu et al
\cite{Liu2009} discussed the New Agegraphic Dark Energy (NADE)
model in the framework of the Brans-Dicke theory with the help of
EoS and deceleration parameters and as a result they showed that
the EoS parameter has a quintom-like behavior for that model and
it indicates the accelerated expansion of the universe. The NADE
in $f(R)$ gravity model has been discussed in \cite{Setare2010}
where it is find that there may exist a phantom-like universe.
Also in \cite{Jamil2010} the NADE model is correspond with
Horava-Lifshitz gravity which indicates that the accelerated
expanding universe is consistent with cosmological observations.
The reconstruction of entropy-corrected Holographic Dark Energy
(ECHDE) in the $f(G)$ gravity also been investigated to explain
the expanding universe with acceleration in \cite{Setare2010b}.
Recently the HDE model in the framework of the $f(G)$ gravity has
been discussed in \cite{Jawad2013} and the different phenomenon
for the accelerating universe are explained.\\

Motivated by the work of Debnath \cite{Debnath2013}, we
reconstructed the Einstein-Aether gravity models from the modified
gravities like $f(T)$, $f(R)$, $f(G)$, $f(R,T)$ and $f(R,G)$ by
assuming a power law solution of the scale factor $(a=a_{0}
t^{n})$ separately. For this purpose, we first briefly explained
the Einstein-Aether gravity theory by modifying the
Einstein-Hilbert action in section II and with the help of the
modified Friedmann equations, we evaluated the effective density
and pressure for Einstein-Aether gravity sector. After that in
Sections III-VI, we have shown the correspondence between
Einstein-Aether gravity and other modified gravities $f(T)$,
$f(R)$, $f(G)$, $f(R,T)$ and $f(R,G)$ and where we have
reconstructed the unknown function $F(K)$ in term of $K$ and
analyzed the nature of $F(K)$ in graphically. In Section VIII, we
analyzed the stability of each reconstructed Einstein-Aether
Gravity model. Finally, we described some cosmological
implications of these reconstructed models in section IX.

\section{\normalsize\bf{Modified Friedmann Equations in Einstein-Aether Gravity Theory}}
Einstein-Aether Gravity Theory is basically a generalization of
General Relativity(GR). The theory describes that the spacetime is
coupled with both a metric and a dynamical unit timelike vector
field named Aether. As it has a preferred reference frame and
hence it violates Lorentz invariant. The action of the
Einstein-Aether gravity theory with the normal Einstein-Hilbert
part action can be written in the form \cite{Jacob,Zlos,Meng1}
\begin{equation}
S=\int d^{4}x\sqrt{-g}\left[\frac{R}{16\pi G}+{\cal L}_{EA}+{\cal
L}_{m} \right]\label{2.1}
\end{equation}

where ${\cal L}_{EA}$ is the vector field Lagrangian density while
${\cal L}_{m}$ denotes the Lagrangian density for all other matter
fields. The Lagrangian density for the Einstein-Aether gravity is
given by \cite{Zlos,Meng1}:
\begin{equation} {\cal L}_{EA}=\frac{M^{2}}{16\pi
G}~F(K)+\frac{1}{16\pi G}~\lambda(A^{a}A_{a}+1)~,\label{2.2}
\end{equation}
\begin{equation}
K=M^{-2}(c_{1}g^{ab}g_{cd}+c_{2}\delta^{a}_{c}\delta^{b}_{d}+
c_{3}\delta^{a}_{d}\delta^{b}_{c})\nabla_{a}A^{c}\nabla_{b}A^{d}~\label{2.3}
\end{equation}
where $c_{i}$ are dimensionless constants, $M$ is the coupling
constant, $\lambda$ is a Lagrangian multiplier and $F(K)$ ia an
arbitrary function of $K$. From (\ref{2.1}), we get the Einstein's
field equations
\begin{equation}
G_{ab}=T_{ab}^{EA}+8\pi G T_{ab}^{m}\label{2.4}
\end{equation}
Here $T_{ab}^{m}$ is the energy momentum tensor for matter field
and $T_{ab}^{EA}$ is the energy momentum tensor for the vector
field and they are respectively given as follows: \cite{Meng1}
\begin{equation}
T_{ab}^{m}=(\rho_{m}+p_{m})u_{a}u_{b}+p_{m}g_{ab}\label{2.5}
\end{equation}
where $\rho_{m}$ and $p_{m}$ are respectively the energy density
and pressure of matter and $u_{a}=(1,0,0,0)$ is the fluid
4-velocity vector and
\begin{equation}
T_{ab}^{EA}=\frac{1}{2}~\nabla_{d}\left[
\left({J_{(a}}^{d}A_{b)}-{J^{d}}_{(a}A_{b)}-J_{(ab)}A^{d}
\right)F' \right]-Y_{(ab)}F'+\frac{1}{2}~g_{ab}M^{2}F+\lambda
A_{a}A_{b}\label{2.6}
\end{equation}
with
\begin{equation}
Y_{ab}=-c_{1}\left[
(\nabla_{d}A_{a})(\nabla^{d}A_{b})-(\nabla_{a}A_{d})(\nabla_{b}A^{d})
\right]\label{2.7}
\end{equation}
where the subscript $(ab)$ means symmetric with respect to the
indices involved and $A^{a}=(1,0,0,0)$ is non-vanishing time-like
unit vector satisfying $A^{a}A_{a}=-1$.\\

 Now we consider the Friedmann-Robertson-Walker (FRW) metric of the
universe as
\begin{equation}
ds^{2}=-dt^{2}+a^{2}(t)\left[\frac{dr^{2}}{1-kr^{2}}+r^{2}\left(d\theta^{2}+sin^{2}\theta
d\phi^{2}\right) \right]\label{2.8}
\end{equation}
where $k~(=0,\pm 1)$ is the curvature scalar and $a(t)$ is the
scale factor. From equation (3), we get
\begin{equation}
K=\frac{3\beta H^{2}}{M^{2}}\label{2.9}
\end{equation}
where $H~(=\frac{\dot{a}}{a})$ is Hubble parameter and the
coefficient $\beta$ is expressed as $\beta=c_1+3c_2+c_3$
\cite{Meng1,Clifton} and it is constant. From eq. (\ref{2.4}), we
get the modified Friedmann equations (for flat universe $k=0$) for
Einstein-Aether gravity as in the following \cite{Zlos,Meng1} form

\begin{equation}
\beta\left(-F'+\frac{F}{2K}\right)H^{2}+H^{2}=\frac{8\pi
G}{3}~\rho \label{2.10}
\end{equation}
and
\begin{equation}
\beta\frac{d}{dt}\left(HF'\right)-2\dot{H}=8\pi G(\rho+p)
\label{2.11}
\end{equation}
Now, if $\rho_{EA}$ and $p_{EA}$ are considered as the effective
energy density and pressure governed by the Einstein-Aether
gravity, then we can write the eqs. (\ref{2.10})\&(\ref{2.11}) in
the following form:
\begin{equation}
H^{2}=\frac{1}{3}\rho +\frac{8\pi G}{3}\rho_{EA}
\end{equation}
and
\begin{equation}
-2\dot{H}=8\pi G(\rho+p)+(\rho_{EA}+p_{EA})
\end{equation}
where the effective energy density and pressure for
Einstein-Aether gravity sector are given by (choosing $8\pi G=1$)
\begin{equation}
\rho_{EA}=3\beta H^{2}\left(F'-\frac{F}{2K} \right)\label{2.12}
\end{equation}
and
\begin{equation}
p_{EA}=-3\beta H^{2}\left(F'-\frac{F}{2K}
\right)-\beta(\dot{H}F'+H\dot{F}')\label{2.13}
\end{equation}

At this point we want to discuss some special cases due to the
power law form of the function $F(K)$. It can be noted that the
modified Friedmann equations (\ref{2.10})\&(\ref{2.11}) can be
rewritten as \cite{Clifton}
\begin{equation}\label{2.10.1}
\left[1-\beta
\sqrt{K}\frac{d}{dK}\left(\frac{F}{\sqrt{K}}\right)\right]H^{2}
=\frac{8\pi G}{3}\rho
\end{equation}
and
\begin{equation}\label{2.11.2}
\frac{d}{dt}\left[-2H+\beta HF_{K}\right]=8\pi G(\rho+p)
\end{equation}
and if we consider $F(K)=\gamma K^{n}$ where $\gamma$ is a
constant then eq. (\ref{2.10.1}) become
\begin{equation}\label{2.11.3}
\left[1+\epsilon\left(\frac{H}{M}\right)^{2(n-1)}\right]
H^{2}=\frac{8\pi G}{3}\rho
\end{equation}
where $\epsilon\equiv \frac{1}{6}(1-2n)\gamma 3^{n}\beta^{n}$ and
also from eq. (\ref{2.11.3}) we have
\begin{equation}\label{2.11.4}
\gamma=\frac{6(\Omega_{m}-1)}{(1-2n)3^n
\beta^n}\left(\frac{M}{H_0}\right)^{2(n-1)}
\end{equation}
where $\Omega_m=8\pi G\rho_0/3H_0^{2}$ and $H_0$ is the present
value of Hubble parameter. Let us now consider some particular
value of $n$: If $n=1$, we have $\epsilon=-~\gamma\beta/2$ and
Newton's constant is rescaled by a factor
$1/(1+\epsilon)$\cite{Carroll} ; If $n=1/2$, then we have the
unchanged Friedmann equation as $\epsilon=0$ and in this scenario
there is no effect on the background cosmology; If $n=0$ we get
the cosmological constant $\Lambda\simeq sign(\gamma)M^2$.
Therefore we can obtain different regimes depending on $n$ for
that special scenario of power law form of the function $F(K)$.

In the following sections, we shall study the correspondence
between Einstein-Aether gravity and other modified gravities like
$f(T)$, $f(R)$, $f(G)$, $f(R,T)$ and $f(R,G)$.

\section{\normalsize\bf{CORRESPONDENCE BETWEEN EINSTEIN-AETHER
AND} $f(T)$ \normalsize\bf{GRAVITIES}} The action for the $f(T)$
gravity is given by \cite{Li2,Myrzakulov,Wu}
\begin{equation}
S_T=\frac{1}{2}\int d^4x
\sqrt{-g}\left[f(T)+{\cal{L}}_m\right]\label{3.14}
\end{equation}
where $T$ is the torsion scalar which can be chosen as $T=-6H^2$,
$f(T)$ is general differentiable function of the torsion and $8\pi
G=c=1$. The modified Friedmann equations for $f(T)$ gravity can be
written as,
\begin{equation}
H^2=\frac{1}{3}(\rho_{m}+\rho_{T})\label{3.15}
\end{equation}
and
\begin{equation}
\dot{H}=-\frac{1}{2}(\rho_{m}+p_{m}+\rho_{T}+p_{T})\label{3.16}
\end{equation}
where $\rho_{T}$ and $p_{T}$ be the energy density and pressure in
$f(T)$ gravity given by
\begin{equation}
\rho_{T}=\frac{1}{2}(2Tf'(T)-f(T)+6H^2)\label{3.17}
\end{equation}
and
\begin{equation}
p_T=-\frac{1}{2}(-8\dot{H}Tf''(T)+(2T-4\dot{H})f'(T)-f(T)+4\dot{H}+6H^2)
\label{3.18}
\end{equation}
where prime and dot denote the derivatives w.r.t. $T$ and $t$
respectively.

Now we make correspondence between Einstein-Aether gravity and
$f(T)$ gravity by equating their energy densities (\ref{2.12}) and
(\ref{3.17}). For this purpose, we assume a simple power law form
of $f(T)=f_{T0}T^{m}$ ($m>0$), so we get the differential equation
for $F(K)$ as in the form:
\begin{equation}
\frac{dF}{dK}-\frac{F}{2K}=A_{T}K^{m-1}\label{3.19}
\end{equation}
where,
\begin{equation}
A_{T}=\frac{(-1)^m}{\beta^m}(1-2m)2^{m-1}M^{2m-2}f_{T0}\label{3.20}
\end{equation}
and from which we have
\begin{equation}
F(K)=\frac{2A_{T}}{2m-1}K^m+B_{T}\sqrt{K}\label{3.21}
\end{equation}
where $B_{T}$ is a constant. We observe that $F(K)$ is analytic
function of $K$. So $F(K)$ can be reconstructed in the framework
of $f(T)$ gravity. Also in figure 1, we have drawn the function
$F(K)$ for different positive values of $m$. For, $m=2$, the
function $F(K)$ first increases to some upper bound and then
decreases. But for $m=3$ and 4, the function $F(K)$ increases
always.

\begin{figure}
\includegraphics[height=2.5in]{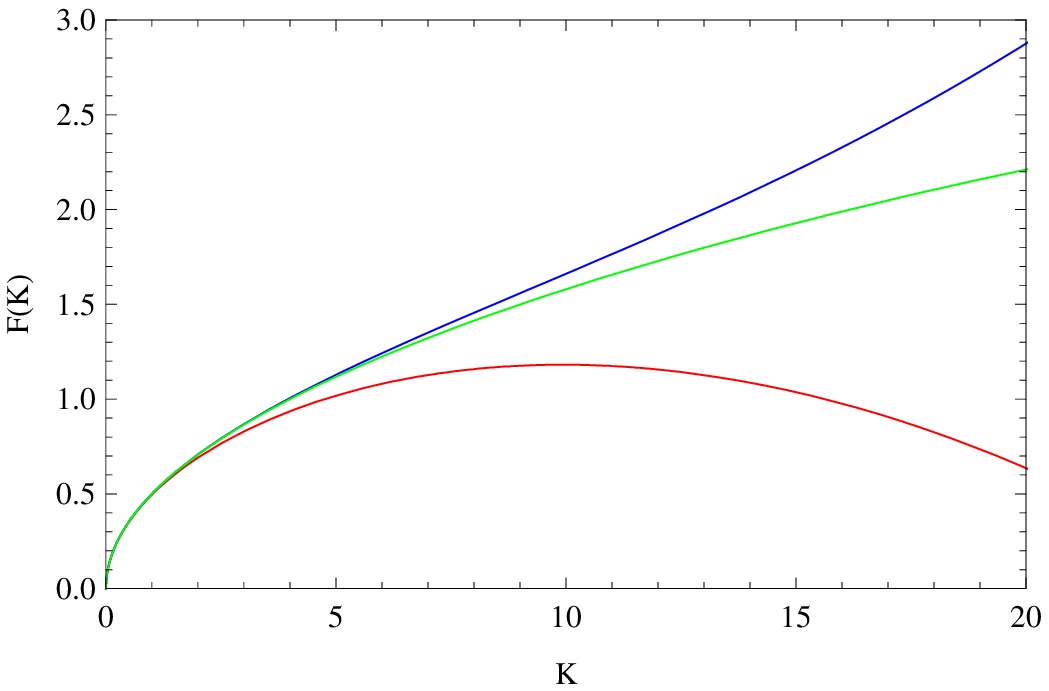}\\
\vspace{2mm}
Fig.1\\
\vspace{4mm} \vspace{.2in} Fig. 1 shows the variation of
reconstructed $F(K)$ against $K$ from $f(T)$ gravity theory where
$M=0.01, \beta=0.1, B=0.5, f_{T0}=1$ and $m=2$ (Red line), $m=3$
(Blue line), $m=4$ (Green line). \vspace{0.2in}
\end{figure}

\section{\normalsize\bf{CORRESPONDENCE BETWEEN EINSTEIN-AETHER
AND} $f(R)$ \normalsize\bf{GRAVITIES}}

The action of $f(R)$ gravity with matter is given in the
four-dimensional flat space-time by
\cite{Nojiri2007a,Li,Nojiri2006},
\begin{equation}
S_R=\int d^4x\sqrt{-g}\left[\frac{R+f(R)}{2}+{\cal{L}}_m\right]
\label{4.22}
\end{equation}
where $R$ is replaced with $f(R)$ to generalized the usual
Einstein-Hilbert action, where $f(R)$ is an analytic function of
$R$ and ${\cal{L}}_m$ is the matter Lagrangian. Also in this flat
space-time ($k=0$), the Ricci scalar is given by
$R=6\dot{H}+12H^2$ and choosing $8\pi G=c=1$.\\

The gravitational field equations in flat ($k=0$) space-time are
given by
\begin{equation}
3H^{2}=\rho_{m}+\rho_{R} \label{4.23}
\end{equation}
and
\begin{equation}
2\dot{H}+3H^{2}=-(p_{m}+p_{R}) \label{4.24}
\end{equation}
where $\rho_{m}$ being the energy density and $p_{m}$ is the
pressure of matter and
\begin{equation}
\rho_{R}=-\frac{1}{2}f(R)+3(\dot{H}+H^{2})f'(R)
-18(4H^{2}\dot{H}+H\ddot{H})f''(R)\label{4.25}
\end{equation}
and
\begin{equation}
p_{R}=\frac{1}{2}f(R)-(\dot{H}+3H^{2})f'(R)+6(8H^{2}\dot{H}+6H\ddot{H}+4\dot{H}^{2}+\dddot{H})f''(R)
+36(\ddot{H}+4H\dot{H})^{2}f'''(R) \label{4.26}
\end{equation}
where prime and dot denote the derivatives w.r.t. $R$ and $t$. Now
we make correspondence between Einstein-Aether gravity theory and
$f(R)$ gravity theory by equating their energy densities
(\ref{2.12}) and (\ref{4.25}). For this purpose, we assume
\begin{equation}
f(R)=f_{R0}R^{\mu} \label{4.27}
\end{equation}
where $f_{R0}$ and $\mu>0$ are constants. Again, we assume the
power law form of the scale factor in the form $a=a_{0}t^{n}$,
where, $a_{0}$ and $n$ are constants. For accelerating phase of
the universe, we must have $n>1$, because the deceleration
parameter $q$ must be negative. Now if we put the power law form
of scale factor and equation (\ref{4.27}) into equations
(\ref{4.22})-(\ref{4.26}), we get the relations between the
unknown constants, which are not important for our present study.
So we get the differential equation in the form
\begin{equation}
\frac{dF}{dK}-\frac{F}{2K}=\frac{A_{R}M^{2\mu-2}}{3^{\mu}\beta^{\mu}n^{2\mu}}K^{\mu-1}
\end{equation}
and from which we have
\begin{equation}
F(K)=\left(\frac{A_{R} M^{2\mu-2}}{3^\mu\beta^\mu n^{2\mu}}\right)
\left(\frac{K^\mu}{\mu-\frac{1}{2}}\right)+B_{R}\sqrt{K}
\end{equation}
where we consider
\begin{equation}
A_{R}=2^{\mu-1} 3^{\mu}n^{\mu} f_{R0}\left(-1+\mu-2 \mu^2+(2+\mu) n\right)(1-2n)^{\mu-1} \\
\end{equation}
and $B_{R}$ is arbitrary constant. We observe that $F(K)$ is
analytic function of $K$ for $\mu\ne \frac{1}{2}$. So $F(K)$ can
be reconstructed in the framework of $f(R)$ gravity. Also in
figure 2, we have drawn the function $F(K)$ for different positive
values of $\mu$. For $\mu=4$, 5 and 6, the function $F(K)$
increases always.

\begin{figure}
\includegraphics[height=2.5in]{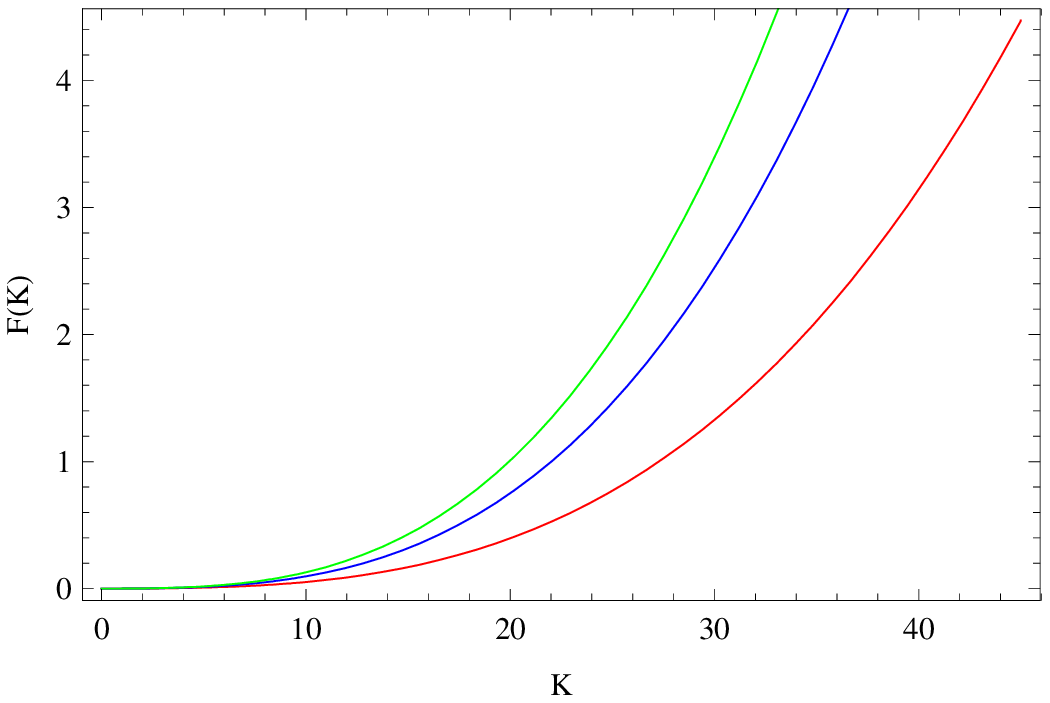}\\
\vspace{.2in}
Fig.2\\
\vspace{.2in} Fig.2 shows the variation of reconstructed $F(K)$
against $K$ from $f(R)$ gravity theory where $n=3, M=0.01,
\beta=0.1, B=0.001, f_{R0}=1$ and $\mu=4$ (Red line), $\mu=5$
(Blue line), $\mu=6$ (Green line). \vspace{0.2in}
\end{figure}

\section{\normalsize\bf{CORRESPONDENCE BETWEEN EINSTEIN-AETHER
AND} $f(G)$ \normalsize\bf{GRAVITIES}}

The action for the $f(G)$ gravity is given by
\cite{Rastkar,Nojiri2005},
\begin{equation}
S_G=\frac{1}{2}\int d^4x
\sqrt{-g}\left[\frac{1}{2}R+f(G)+{\cal{L}}_m\right]\label{5.31}
\end{equation}
where $R$ is the Ricci scalar curvature and ${\cal L}_{m}$ is the
matter Lagrangian. Here $f(G)$ is an arbitrary differentiable
function of $G$ and $G$ is generally defined as
$G=R_{\alpha\beta\gamma\delta}R^{\alpha\beta\gamma\delta}-4R_{\alpha\beta}R^{\alpha\beta}+R^{2}$,
where $R_{\alpha\beta}$ and $R_{\alpha\beta\gamma\delta}$ are
Ricci and Riemann curvature tensors, respectively. Now, the
variation of the action $S$, given by eq. (\ref{5.31}) with
respect to the metric tensor $g_{\alpha\beta}$ is given as follows
\cite{Rastkar,Nojiri2005prd}:
\begin{eqnarray*}
0=\frac{1}{2}\left(-R^{\alpha\beta}
+\frac{1}{2}g^{\alpha\beta}R\right)+T^{\alpha\beta} +
\frac{1}{2}g^{\alpha\beta}f(G)-2f'(G)RR^{\alpha\beta}
+4f'(G)R_{\gamma}^{\alpha} R^{\beta\gamma}
-2f'(G)R^{\alpha\gamma\delta\tau}R_{\gamma\delta\tau}^{\beta}
\end{eqnarray*}
\begin{eqnarray*}
+4f'(G)R^{\alpha\gamma\delta\beta}R_{\gamma\delta}
+2(\nabla^{\alpha}\nabla^{\beta}f'(G))R
-2g^{\alpha\beta}(\nabla^{2}f'(G))R
-4(\nabla_{\gamma}\nabla^{\alpha}f'(G))R^{\beta\gamma}
\end{eqnarray*}
\begin{equation}
-4(\nabla_{\gamma}\nabla^{\beta}f'(G))R^{\alpha\gamma}
+4(\nabla^{2}f'(G))R^{\alpha\beta}
+4g^{\alpha\beta}(\nabla_{\gamma}
\nabla_{\delta}f'(G))R^{\gamma\delta}
-4(\nabla_{\gamma}\nabla_{\delta}f'(G))R^{\alpha\gamma\beta\delta}\label{5.32}
\end{equation}
Now the metric of flat Friedmann-Robertson-Walker(FRW) universe is
defined as
\begin{equation}
ds^{2}=-dt^{2}+a(t)^{2}\sum_{i=1}^{3}(dx^{i})^{2}\label{5.33}
\end{equation}
where $a(t)$ is the scale factor of the cosmic time $t$. From Eq.
(\ref{5.33}) with the help of Eq.(\ref{5.32}) we have
\begin{equation}
0=-f(G)+Gf'(G)-24H^{3}\dot{G}f''(G)-6H^{2}+2\rho_{m}
\end{equation}
and
\begin{equation}
0=f(G)-Gf'(G)+16H^{3}\dot{G}f''(G)+
16H\dot{H}\dot{G}f''(G)+8H^{2}(\ddot{G}f''(G)+\dot{G}^{2}f'''(G))
+2(2\dot{H}+3H^{2})+2p_{m}
\end{equation}

Therefore the modified Fridmann equations can be rewritten as,
\begin{equation}
3H^{2}=\rho_{m}+\rho_{G}
\end{equation}
and
\begin{equation}
-(2\dot{H}+3H^{2})= p_{m}+p_{G}
\end{equation}
where
\begin{equation}
\rho_{G}=\frac{1}{2}\left[-f(G)
+Gf'(G)-24H^{3}\dot{G}f''(G)\right]\label{5.38}
\end{equation}
and
\begin{equation}
p_{G}=\frac{1}{2}\left[f(G)-Gf'(G)+16H^{3}\dot{G}f''(G)+
16H\dot{H}\dot{G}f''(G)+8H^{2}(\ddot{G}f''(G)+\dot{G}^{2}f'''(G))\right]
\end{equation}
where prime and dot denote the derivatives w.r.t. $G$ and $t$
respectively and the Gauss-Bonnet invariant
$G=24H^{2}(H^{2}+\dot{H})$.\\

Now we make correspondence between Einstein-Aether gravity theory
and $f(G)$ gravity theory by equating their energy densities
(\ref{2.12}) and (\ref{5.38}). For this purpose, we assume
\begin{equation}
f(G)=f_{G0}G^{\nu}
\end{equation}
where $f_{G0}$ and $\nu>0$ are constants. Again, we assume the
power law form of the scale factor in the form $a=a_{0}t^{n}$,
where, $a_{0}$ and $n$ are constants. So we get the differential
equation in the form
\begin{equation}
\frac{dF}{dK}-\frac{F}{2K}=A_{G} K^{\nu-1}
\end{equation}
and from which we have the solution
\begin{equation}
F(K)=\frac{2 A_{G}}{4\nu-1}K^{2\nu}+B_{G}\sqrt{K}
\end{equation}
where we consider
\begin{equation}
A_{G}=\frac{\left(\frac{8}{3}\right)^\nu f_{G0} (-1+\nu) (-1+4
\nu+n) \left(\frac{ M^4 (-1+n)}{n \beta ^2}\right)^\nu}{2 M^2
(-1+n)}
\end{equation}
and $B_{G}$ is arbitrary constant. We observe that $F(K)$ is
analytic function of $K$ for $\nu\ne\frac{1}{4}$. So $F(K)$ can be
reconstructed in the framework of $f(G)$ gravity. Also in figure
3, we have drawn the function $F(K)$ for different positive values
of $\nu$. For, $\nu=2$, the function $F(K)$ first increases to
some upper bound and then decreases. But for $\nu=3$ and 4, the
function $F(K)$ increases always.

\begin{figure}
\includegraphics[height=2.5in]{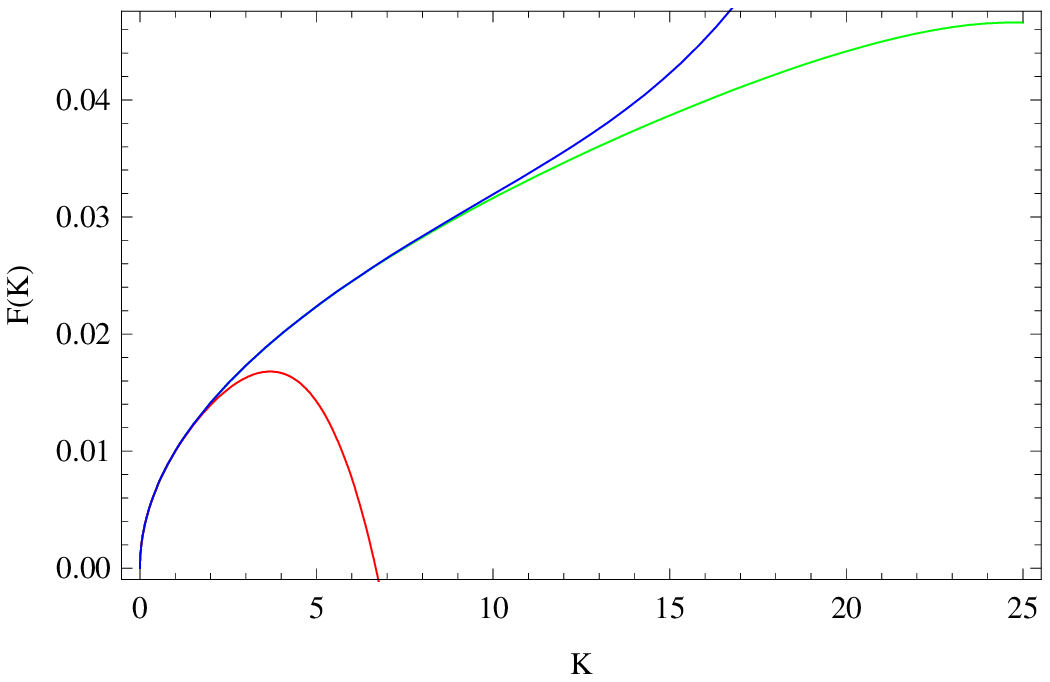}\\
\vspace{2mm}
~~~~~Fig.3\\
\vspace{.2in}Fig.3 shows the variation of reconstructed $F(K)$
against $K$ from $f(G)$ gravity theory where $n=0.1, M=0.01,
\beta=0.1, B_G=0.01$ and $\nu=2$ (Red line), $\nu=3$ (Blue line),
$\nu=4$ (Green line). \vspace{0.2in}
\end{figure}

\section{\normalsize\bf{CORRESPONDENCE BETWEEN EINSTEIN-AETHER
AND} $f(R,T)$ \normalsize\bf{GRAVITIES}}

The action for the $f(R,T)$ gravity is given by
\cite{Harko2011,Jamil2012,Sharif2012},
\begin{equation}
S=\frac{1}{2}\int d^4x \sqrt{-g}\left[\frac{f(R,T)}{16\pi
G}+{\cal{L}}_m\right]
\end{equation}
where ${\cal{L}}_m$ denotes the matter contents of the universe.
In FRW background, the gravitational field equation are given by
\begin{equation}
3H^{2}=8\pi G_{eff}(\rho_{m}+\rho_{RT})
\end{equation}
\begin{equation}
\dot{H}=-4\pi G_{eff}(\rho_{m}+p_{m}+\rho_{RT}+p_{RT})
\end{equation}
where
\begin{equation}
\rho_{RT}=3H^{2}\frac{f_{T}-f_{R}+1}{1+f_{T}}+\frac{
\frac{1}{2}(Rf_{R}-f)-3H\dot{R}f_{RR}}{1+f_{T}}\label{6.47}
\end{equation}
\begin{equation}
p_{RT}=\frac{1}{1+f_{T}}\left[-\frac{1}{2}(Rf_{R}-f)+
2H\dot{R}f_{RR}+\ddot{R}f_{RR}+\dot{R}^{2}f_{RRR}\right]
\end{equation}
and
\begin{equation}
G_{eff}=\frac{1}{f_{R}(R,T)}\left(G+\frac{f_{T}(R,T)}{8\pi}\right)
\end{equation}
is the effective gravitational matter dependent coupling in
$f(R,T)$ gravity Here $\rho_{RT}$ and $p_{RT}$ are the energy
density and pressure of dark energy and choosing $8\pi G=c=1$.
Here, the prime denotes the non-equilibrium description of the
field equations. Now we make correspondence between
Einstein-Aether gravity theory and $f(R,T)$ gravity theory by
equating their energy densities (\ref{2.12}) and (\ref{6.47}). For
this purpose, we assume
\begin{equation}
f(R,T)=d_{1}R^{\mu_{1}}+d_{2}T^{\mu_{2}}
\end{equation}
where $d_{1}$, $d_{2}$, $\mu_{1}>0$ and $\mu_{2}>0$ are constants.
Again, we assume the power law form of the scale factor in the
form $a=a_{0}t^{n}$, where, $a_{0}$ and $n$ are constants. So we
get the differential equation in the form

\begin{eqnarray*}
\frac{dF}{dK}-\frac{F}{2K}=\frac{1}{\beta }+\left(\frac{d_2\mu
_2}{\beta }\left(-6n^2\right)^{\mu _2-1}-\frac{d_1\mu _1}{\beta
}(6n(6n-1))^{1-\mu _1}-\frac{d_2}{6\beta
 n^2}\left(-6n^2\right)^{\mu _2}\right)
\left(\frac{3\beta  n^2}{M^2}\right)^{1-\mu _2} K^{\mu _2-1}
\end{eqnarray*}
\begin{equation}
+ \left(\frac{\left(\mu _1-1\right)}{6\beta  n^2}(6n(6n-1))^{\mu
_1}+\frac{12d_1\mu _1(2n-1)\left(\mu _1-1\right)}{\beta
}(6n(6n-1))^{\mu _1-2}\right) \left(\frac{3\beta
n^2}{M^2}\right)^{1-\mu _1} K^{\mu _1-1}
\end{equation}
and from which we have (for $\mu_2=2$)
\begin{eqnarray*}
F(K)=\frac{B_{RT}}{\sqrt{K}}-\frac{1}{8 \sqrt{K} M^2 (-1+2 n)
\beta d_2^{3/2} \left(M+2 M \mu _1\right)} \left((-1+2 n) \beta
^{3/2} tanh^{-1}\left[\frac{2 \sqrt{K} M \sqrt{d_2}}{\sqrt{\beta
}}\right] \left(1+2 \mu _1\right)-\right.\\2 \sqrt{K} M (-1+2 n)
\beta \sqrt{d_2} \left(1+2 \mu _1\right)- 8 \sqrt{K} M d_2^{3/2}
\left(K M^2 (-1+2 n) \left(1+2 \mu _1\right)+2^{\mu _1}
\left(\frac{K M^2 (-1+2 n)}{n \beta }\right)^{\mu _1} \beta
\right.
\end{eqnarray*}
\begin{equation}
\times\left.\left._{2}F_{1}\left[\frac{1}{2}+\mu_1,1,\frac{3}{2}+\mu_1,\frac{4
K M^2 d_2}{\beta }\right] d_1 \left(1-2 n+(-3+n) \mu _1+2 \mu
_1^2\right)\right)\right)
\end{equation}

and $B_{RT}$ is arbitrary constant. Now $F(K)$ can be
reconstructed in the framework of $f(R,T)$ gravity. Also in figure
4, we have drawn the function $F(K)$ for different positive values
of $d_{1}$, $d_{2}$, $\mu_{1}$ and $\mu_{2}$ and we see that the
function $F(K)$ increases always.

\begin{figure}
\includegraphics[height=2.5in]{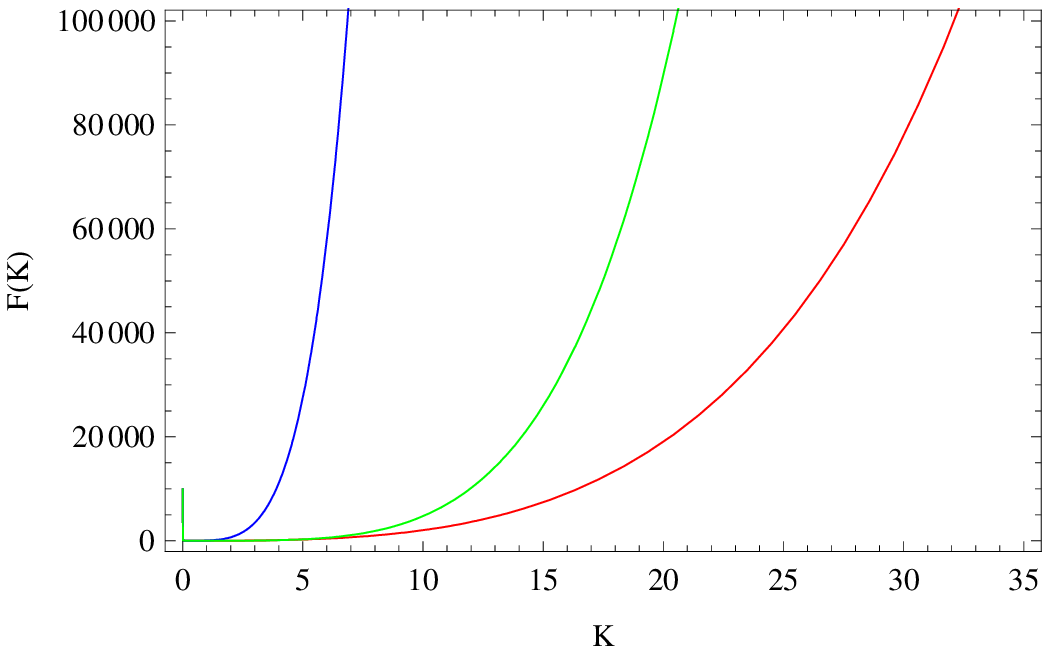}\\
\vspace{2mm}
~~~~~~~Fig.4\\
\vspace{.2in} Fig.4 shows the variation of reconstructed $F(K)$
against $K$ from $f(R,T)$ gravity theory where $n=5, M=2, \beta=1,
B_{RT}=10$ and $\mu_1=3,\mu_2=2,d_1=0.01,d_2=0.001$ (Red line),
$\mu_1=4,\mu_2=2,d_1=0.01,d_2=0.001$ (Blue line),
$\mu_1=4,\mu_2=2,d_1=0.0001,d_2=0.001$ (Green line).
\vspace{0.2in}
\end{figure}

\section{\normalsize\bf{CORRESPONDENCE BETWEEN EINSTEIN-AETHER
AND} $f(R,G)$ \normalsize\bf{GRAVITIES}}

The action of $f(R,G)$ gravity is given by
\cite{Bamba2010,Myrzakulov2013},
\begin{equation}
S_{RG}=\int d^4x\sqrt{-g}[\frac{1}{2}f(R,G)+{\cal{L}}_m]
\end{equation}

In the case of flat FRW Universe, the Einstein's first field
equation is
\begin{equation}
3H^{2}=\rho_{m}+\rho_{RG}
\end{equation}
The contribution of energy density for $f(R,G)$ gravity is given
by
\begin{equation}
\rho_{RG}=3H^{2}+\frac{1}{2}(Gf_{G}-f(R,G)-24H^{3}f_{Gt})
+3(\dot{H}+H^{2})f_{R}-3Hf_{Rt}\label{7.55}
\end{equation}
where $f(R,G)$ is a more general kind of modified Gauss-Bonnet
gravity and suffix denote the partial derivatives. Now we make
correspondence between Einstein-Aether gravity theory and $f(R,G)$
gravity theory by equating their energy densities (\ref{2.12}) and
(\ref{7.55}). For this purpose, we assume
\begin{equation}
f(R,G)=d_{3}R^{\mu_{3}}+d_{4}G^{\mu_{4}}
\end{equation}
where $d_{3}$, $d_{4}$, $\mu_{3}>0$ and $\mu_{4}>0$ are constants.
Again, we assume the power law form of the scale factor in the
form $a=a_{0}t^{n}$, where, $a_{0}$ and $n$ are constants. So we
get the differential equation in the form

\begin{eqnarray*}
\frac{dF}{dK}-\frac{F}{2K}=\frac{1}{2 K M^2}\left(\frac{2 K
M^2}{\beta }+\frac{2^{\mu _3} \left(\frac{K M^2 (-1+2 n)}{n \beta
}\right)^{\mu _3} d_3 \left(1-2 n+\mu _3 \left(-3+n+2 \mu
_3\right)\right)}{-1+2 n}\right.
\end{eqnarray*}
\begin{equation}
\left.~~~~~~~~~~~~~~~~~~~~~~~~~~~~~~~~~~~~~~~~~~~~~~~~~
+\frac{\left(\frac{8}{3}\right)^{\mu _4} \left(\frac{K^2 M^4
(-1+n)}{n \beta ^2}\right)^{\mu _4} d_4 \left(-1+\mu _4\right)
\left(-1+n+4 \mu _4\right)}{-1+n}\right)
\end{equation}
and from which we have
\begin{eqnarray*}
F(K)=\frac{B_{RG}}{\sqrt{K}}+\frac{3^{-1-\mu _4}}{M^2 \left(1-3
n+2 n^2\right) \beta} \left(2\times3^{\mu _4} K M^2 \left(1-3 n+2
n^2\right)\right.~~~~~~~~~~~~~~~~~~~~~~~~~~~~~~~~~~~~~~~~~~~~~~~~~~~~\\
\left.+\frac{2^{\mu _3} 3^{1+\mu _4} (-1+n) \left(\frac{K M^2
(-1+2 n)}{n \beta }\right)^{\mu _3} \beta \text{  }d_3 \left(1-2
n+(-3+n) \mu _3+2 \mu_3^2\right)}{1+2 \mu _3}-\frac{3}{1+4 \mu _4}
\left(\frac{K^{2} M^{4} (-1+n)}{n \beta ^2}\right)^{\mu _4}\right.
\end{eqnarray*}
\begin{equation}
\left.\times\beta d_{4}\left(8^{\mu _4}-3 8^{\mu _4} n+2^{1+3 \mu
_4} n^2+\left(-5 8^{\mu _4}+11 8^{\mu _4} n-2^{1+3 \mu _4}
n^2\right) \mu _4-2^{2+3 \mu _4} (-1+2 n) \mu _4^2\right)\right)
\end{equation}
and $B_{RG}$ is arbitrary constant. Now $F(K)$ can be
reconstructed in the framework of $f(R,G)$ gravity. Also in figure
5, we have drawn the function $F(K)$ for different positive values
of $d_{3}$, $d_{4}$, $\mu_{3}$ and $\mu_{4}$ and we see that the
function $F(K)$ increases always.

\begin{figure}
\includegraphics[height=2.5in]{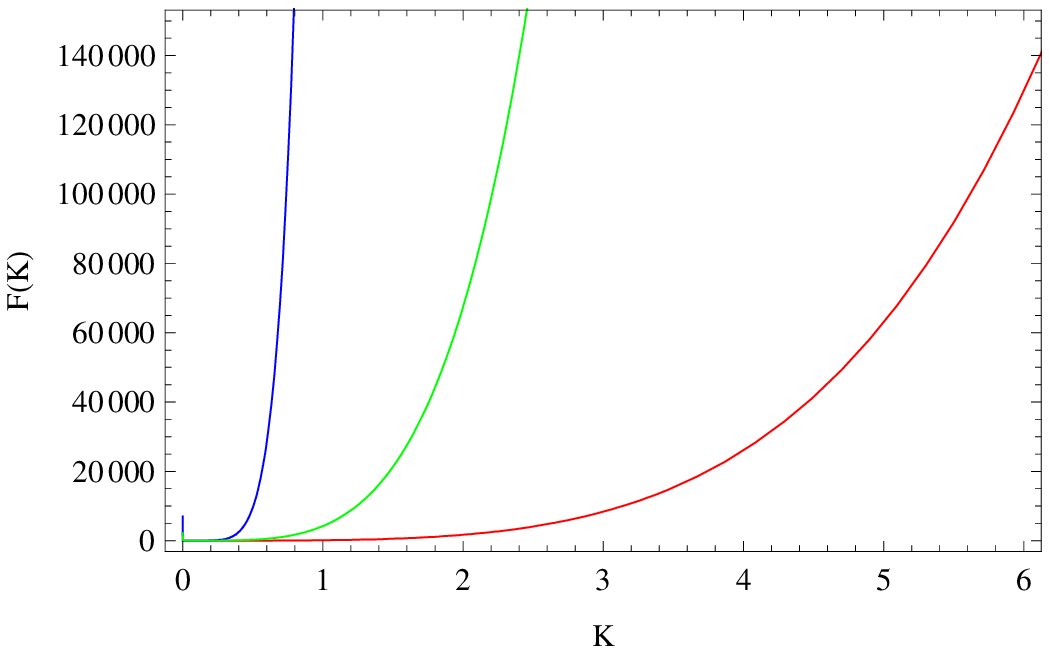}\\
\vspace{2mm}
~~~~Fig.5\\
\vspace{.2in} Fig.5 shows the variation of reconstructed $F(K)$
against $K$ from $f(R,G)$ gravity theory where $n=5, M=2,
\beta=0.1, B_{RG}=1$ and $\mu_3=3,\mu_4=2,d_3=0.0001,d_4=0.0001$
(Red line), $\mu_3=3,\mu_4=3,d_3=0.0001,d_4=0.0001$ (Blue line),
$\mu_3=5,\mu_4=2,d_3=0.0001,d_4=0.0001$ (Green line).
\vspace{0.2in}
\end{figure}

\section{\normalsize\bf{Stability Analysis of Reconstructed Einstein-Aether Gravity Models }}

In this section, we have examined the stability of each
reconstructed Einstein-Aether gravity models from $f(T)$, $f(R)$,
$f(G)$, $f(R,T)$ and $f(R,G)$. In this connection, the most
important quantity is that the square speed of sound which is
denoted as $v_s^2$ and defined as a ratio of the effective
pressure and energy densities i.e.,
$v_s^2=\frac{\dot{p}}{\dot{\rho}}$. The sign of $v_s^2$ plays a
very vital role for stability analysis of a background evolution
of cosmic models. It is well known that the model is stable if
$v_s^2>0$ and if $v_s^2<0$ implies that the model is classically
unstable\cite{Myung}. In 2008, Kim et al. \cite{Kim} found that
$v_s^2$ for agegraphic DE is always negative which leads to
classically instability of that model. Recently many researchers
are using this methods to analyzed the stability of that models on
which they worked and of which some authors
\cite{Jawad,Ebrahimi,Sharif,Setare} have reached to a conclusion
that HDE, ADE, NADE, Chaplygin gas, holographic Chaplygin,
holographic $f(T)$, holographic $f(G)$, new agegraphic $f(T)$, new
agegraphic $f(G)$ models are classically unstable because square
speed of sound is negative i.e., $v_s^2<0$  throughout the
evolution of the universe. Here we consider
\begin{equation}
 v_s^2=\frac{\dot{p}_{EA}}{\dot{\rho}_{EA}}
\end{equation}
and plot $v_s^2$ versus $t$ by taking the power-law scale factor
in the each reconstructed Einstein-Aether Gravity model and fig.
6-10 show the variation of $v_s^2$ with $t$ for those above cases.
In the reconstructed Einstein-Aether gravity model from $f(T)$,
$f(R)$ and $f(G)$ model we observed (Figs. 6-8) that the square
speed of sound remains positive for the present and future epoch,
that implies reconstructed Einstein-Aether Gravity model from
those scenarios with power-law scale factor are classically
stable. Whereas in the case of $f(R,T)$, we found that $v_s^2$
remains negative for the present and future epoch which means
reconstructed Einstein-Aether Gravity model from $f(R,T)$ scenario
with power-law scale factor is classically unstable. But when we
considered the $f(R,G)$ model, we shown that $v_s^2$ remains
positive as well as negative for a certain period of time $t$,
which entailed that the Einstein-Aether Gravity model
reconstructed from $f(R,G)$ is classically stable for a certain
period of time and unstable for remaining epoch.

\begin{figure}
\includegraphics[height=2.2in]{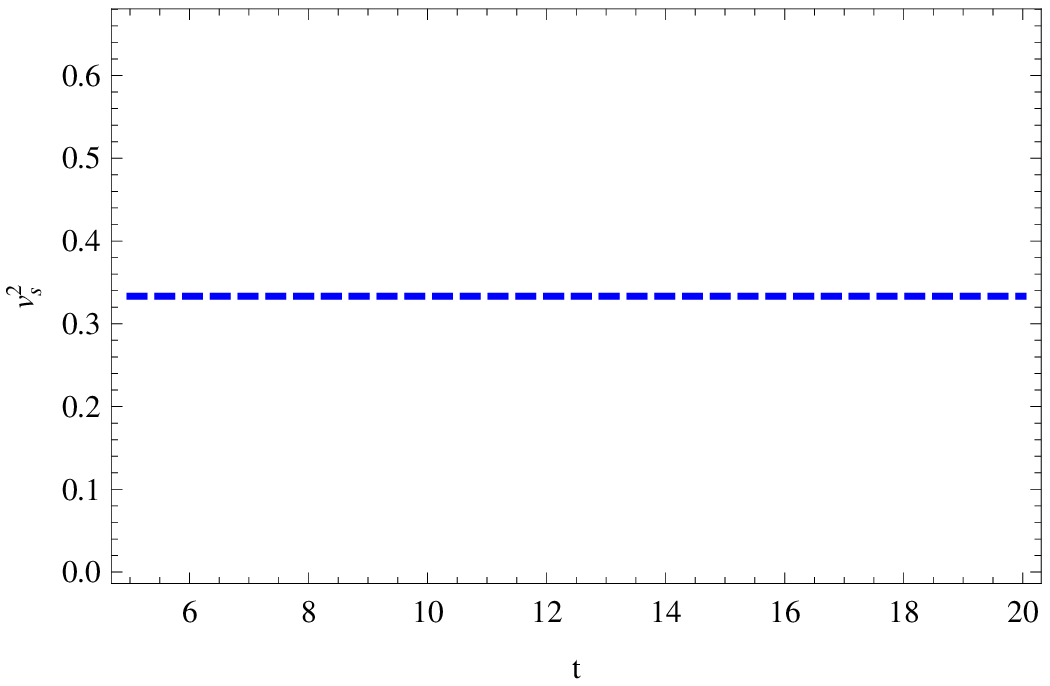}~~
\includegraphics[height=2.2in]{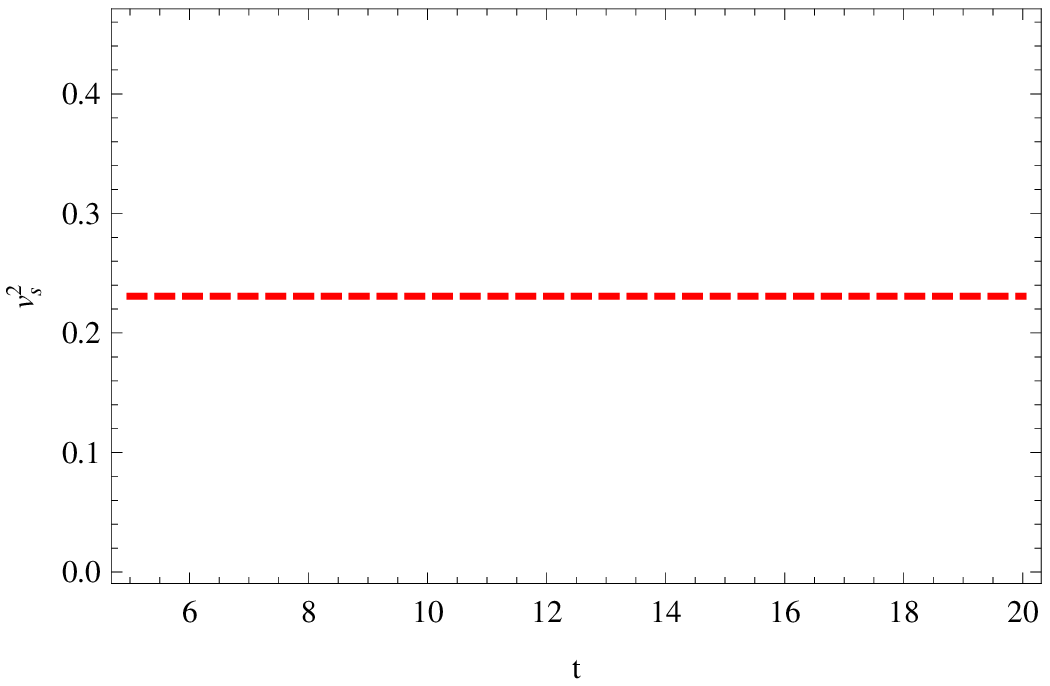}\\
\vspace{2mm}
~~~~~~~~~~Fig.6~~~~~~~~~~~~~~~~~~~~~~~~~~~~~~~~~~~~~~~~~~~~~~~~~~~~~~~~~~~~~~~~~~~~~~~~~~~~Fig.7\\
\vspace{6mm}
\includegraphics[height=2.2in]{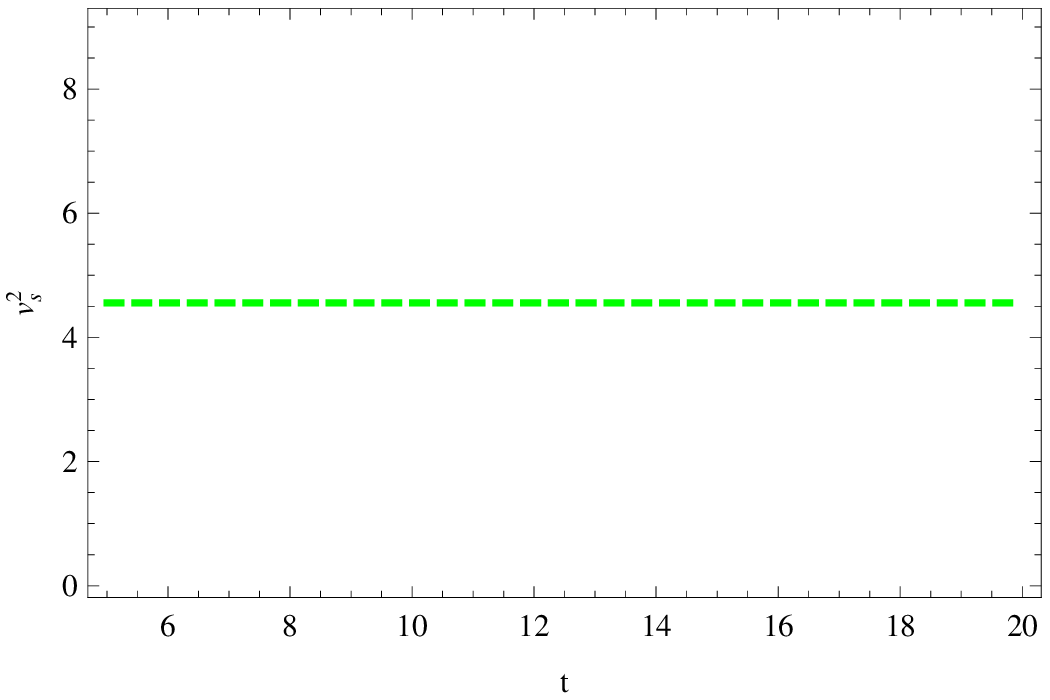}\\
\vspace{2mm}
~~~~~~~~~~~~~~~~~~~~~~~~~~~~~~~~~~~~~~~~~~~~~~~~~~~~~~Fig.8~~~~~~~~~~~~~~~~~~~~~~~~~~~~~~~~~~~~~~~~~~~~~~\\
\vspace{4mm} \vspace{.2in} Figs. 6-8 show the variations of
$v_s^2$ with $t$ for the reconstructed Einstein-Aether theory from
$f(T)$ gravity theory (blue, dashed line), from $f(R)$ gravity
theory (red, dashed line), from $f(G)$ gravity theory (green,
dashed line) respectively. \vspace{0.2in}
\end{figure}

\begin{figure}
\includegraphics[height=2.2in]{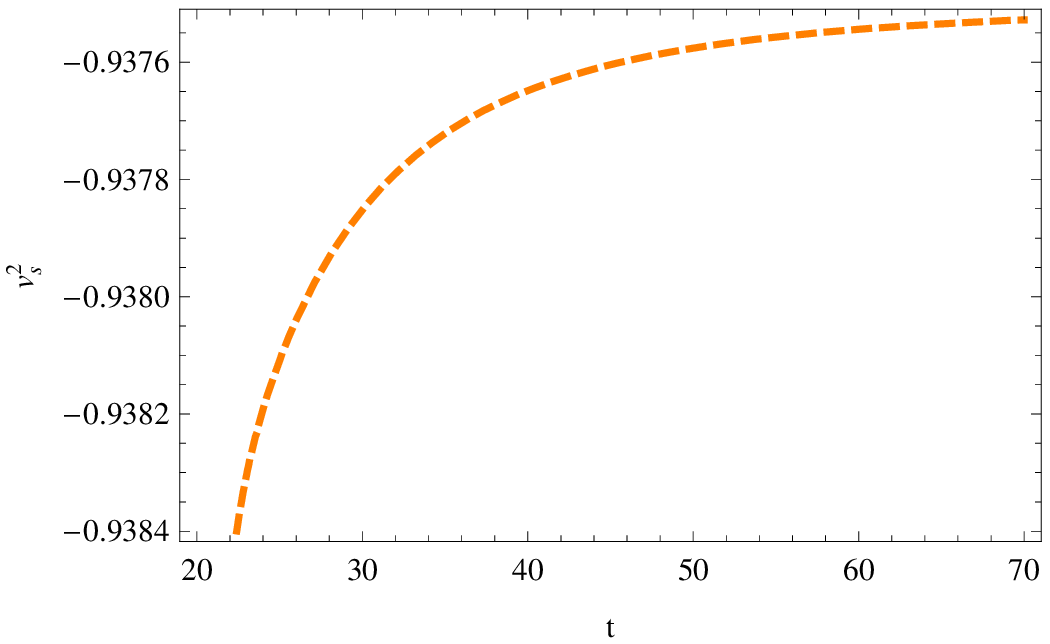}~~
\includegraphics[height=2.2in]{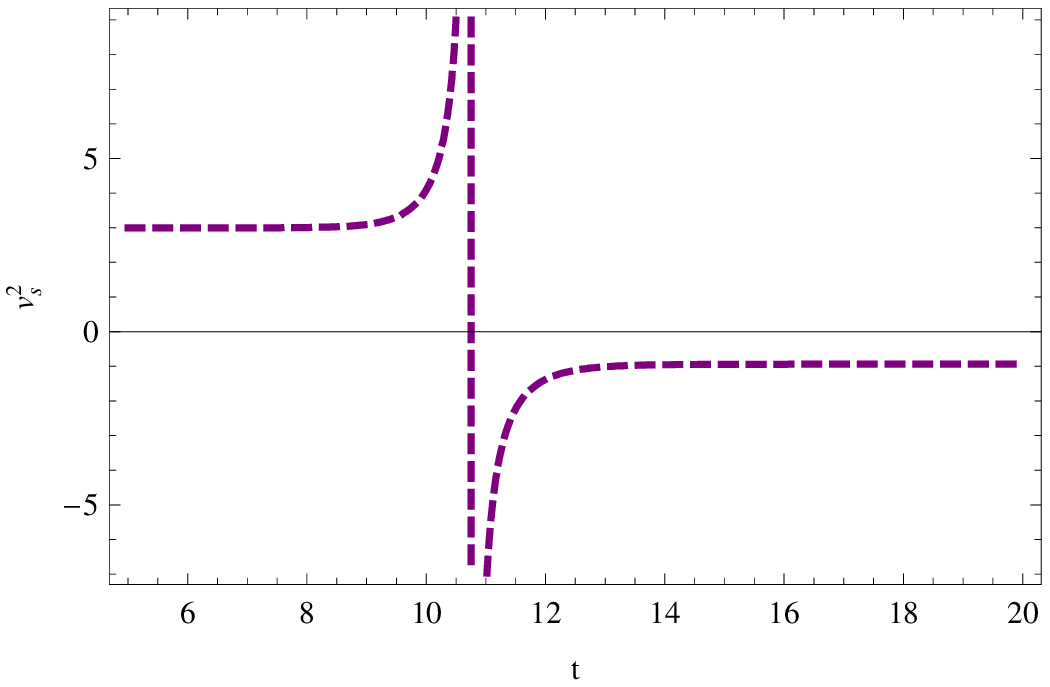}\\
\vspace{2mm}
~~~~~~~~~~~~Fig.9~~~~~~~~~~~~~~~~~~~~~~~~~~~~~~~~~~~~~~~~~~~~~~~~~~~~~~~~~~~~~~~~~~~~~~~~~Fig.10\\
\vspace{4mm} \vspace{.2in} Figs. 9-10 show the variations of
$v_s^2$ with $t$ for the reconstructed Einstein-Aether theory from
$f(R,T)$ gravity theory (orange, dashed line), from $f(R,G)$
gravity theory (purple, dashed line) respectively. \vspace{0.2in}
\end{figure}

\section{\normalsize\bf{Discussions and Concluding Remarks}}
We briefly describe one of the modified gravity named as
Einstein-Aether gravity. In FRW model, the Friedmann equations for
this gravity have been discussed. We find the effective energy
density and pressure for Einstein-Aether gravity. We have analyzed
the correspondence between Einstein-Aether gravity and other
modified gravities like $f(T)$, $f(R)$, $f(G)$, $f(R,T)$ and
$f(R,G)$ gravities by comparing their energy densities. For this
purpose, we have chosen the scale factor in power law form. Next
we have found the free function $F(K)$ for Einstein-Aether gravity
(where $K$ is proportional to $H^{2}$) in term of $K$ in the
framework of the above modified gravity models separately. Thus
the reconstruction is possible for these models. For $f(T)$,
$f(R)$ and $f(G)$ models, we have found the function $F(K)$ is
analytic function in term of $K$. The nature of $F(K)$ vs $K$ have
been shown graphically for every cases. For $f(T)$ gravity model,
we have drawn the function $F(K)$ for different positive values of
$m$ in figure 1. For, $m=2$, the function $F(K)$ first increases
to some upper bound and then decreases. But for $m=3$ and 4, the
function $F(K)$ increases always. For $f(R)$ gravity model, we
have drawn the function $F(K)$ for different positive values of
$\mu$ in figure 2. For $\mu=4$, 5 and 6, the function $F(K)$
always increases. For $f(G)$ gravity model, we have drawn the
function $F(K)$ for different positive values of $\nu$ in figure
3. For, $\nu=2$, the function $F(K)$ first increases to some upper
bound and then decreases. But for $\nu=3$ and 4, the function
$F(K)$ also increases. In figures 4 and 5, we have drawn the
function $F(K)$ for different positive values of constants in our
$f(R,T)$ and $f(R,G)$ gravity models respectively and we see that
the function $F(K)$ increases for these models. Also we checked
the stability of reconstructed Einstein-Aether Gravity model from
$f(T)$, $f(R)$, $f(G)$, $f(R,T)$ and $f(R,G)$ models through the
square speed of sound $v_s^2$ and fig. 6-10 show the variation of
$v_s^2$ with $t$ for those cases. Finally, as results, we shown
that Einstein-Aether Gravity model is classically stable when it
is reconstructed from $f(T)$, $f(R)$ and $f(G)$ models and
classically unstable when it is reconstructed from $f(R,T)$ model
and classically stable for a certain period of time as well as
classically unstable for remaining epoch when it is reconstructed
from $f(R,G)$ model.\\\\\\

{\bf Acknowledgement:}\\

The authors are thankful to IUCAA, Pune, India for warm
hospitality where part of the work was carried out.\\

\end{document}